\documentclass[twocolumn,prb,floatfix,superscriptaddress]{revtex4}

\usepackage{amssymb}
\usepackage{bm}
\usepackage[usenames, dvipsnames]{color}
\usepackage{mathrsfs}
\usepackage{graphicx}
\usepackage{amsmath}
\usepackage{enumitem}
\usepackage{natbib}
\usepackage{float}
\usepackage{verbatim}
\usepackage{color}

\newcommand{\beq}{\begin{equation}}
\newcommand{\eeq}{\end{equation}}
\newcommand{\beqr}{\begin{eqnarray}}
\newcommand{\eeqr}{\end{eqnarray}}
\usepackage{color}

\begin{document}

\title{Entropy production in a photovoltaic cell}
\author{Mohammad H. Ansari}
\affiliation{ Peter Gr\"unberg Institut (PGI-2), Forschungszentrum J\"ulich, D-52425 J\"ulich, Germany}
\affiliation{J\"ulich-Aachen Research Alliance Institute (JARA), Fundamentals of Future Information Technologies, D-52425 J\"ulich, Germany}
\date{\today}
\begin{abstract}
    
We evaluate entropy production in a photovoltaic cell that is modeled by four electronic levels resonantly coupled to thermally populated field modes at different temperatures. We use a formalism recently proposed, the so-called multiple parallel worlds, to consistently address the nonlinearity of entropy in terms of density matrix. Our result shows that entropy production is the difference between two flows: a semiclassical flow that linearly depends on occupational probabilities, and another flow that depends nonlinearly on quantum coherence and has no semiclassical analog. We show that entropy production in the cells depends on environmentally induced decoherence time and energy detuning. We characterize regimes where reversal flow of information takes place from a cold to hot bath. Interestingly, we identify a lower bound on entropy production, which sets limitations on the statistics of dissipated heat in the cells.
 \end{abstract}

\maketitle


\section{Introduction}
\label{intro}

In the past decade a number of  physical quantities, such as charge and spin,  have been  accurately measured in quantum systems\cite{{Horodecki},{vidal},{dur}}, and these measurements have found practical applications in superfast computation and supersecure communication \cite{{clarke},{wendin}}. More recently,  in making use of superconducting qubits and transport by tunneling \cite{{wilhelm},{bal}}, heat dissipation has been  measured in quantum devices \cite{{seifert},{toyabe},{esposito}}, although yet in the absence of full quantum features \cite{{koski},{pekola}}. Industrial photocell technology has reached a saturation in the efficiency of converting solar energy to electricity, and by recent quantum control of heat flow these cells achieve higher efficiencies \cite{schullypc}. 

All these indicate how important is to understand a \emph{consistent} theory for quantum thermodynamics.  The ultimate goal of such a theory is to introduce possible correspondences between information and physical quantities. These correspondences are the textbook laws of thermodynamics in deterministic classical systems. In stochastic systems\cite{{Maruyama},{Dorfmana},{sambur},{cho},{schaller}}, however, they are a set of relations such as the Jarzynski inequality \cite{jar} and the Crook fluctuation theorem \cite{crooks}.  In quantum devices the existence of universal correspondences between information and physics are a subject of research. Here we study the correspondence for entropy as an informational measure.

Entropy is one of the central quantities, whose consistent evaluation in quantum theory is obscure due to its nonlinear dependence on density matrix; $S=-k_B \textup{Tr} \hat{\rho} \ln \hat{\rho}$, with $\rho$ being density matrix.  This quantity is one of the fundamental characteristics for quantifying many-body correlations and proved useful in critical phenomena, quantum quenches, topologically-ordered states, strongly correlated systems, etc.\cite{Amico} In quantum information theory entropy helps to identify sources of fidelity loss.\cite{wilde}  Standard time-evolution formalisms in open quantum systems \cite{{nazarovbook},{Keldysh}} that allow to compute density matrices at different times, are useless in evaluating  entropy due to its nonlinear dependence on  density matrix.\cite{{Bekjan},{Ando}} Recently some progresses have been made to consistently evaluate it in the weak coupling regime\cite{nazarov11}, using the so-called \emph{extended Keldysh technique on multiple parallel worlds} \cite{{ansari3},{ansari1},{ansari2}}. In this terminology the system of  interest and whatever is coupled to it make a \emph{world}. This evaluation beyond perturbation theory is still an open problem. \cite{ansariST}   
 
 We previously calculated entropy production in  simple examples of quantum heat engines.\cite{{ansari3},{ansari1},{ansari2},{ansari4}}   A quantum heat engines (QHE) is a small quantum system with a number of energy levels that are coupled to several heat reservoirs. These devices are known for converting incoherent photons of thermal environments into coherent emissions\cite{schully}. Our results for simple QHEs showed that entropy flow has two parts: 1) an incoherent part, which linearly depends on the quantum system density matrix, and 2) a coherent part, which is nonlinear. Given that both parts depends on density matrix the reason for using this terminology, in the first place of Ref. [\onlinecite{ansari1}], has been that the nonlinear part is independent of occupation probabilities and only depends on quantum coherence element of density matrix as a result of the coherent drive.   Interestingly this part  has no semiclassical analogue.  Separately we showed in Ref. [\onlinecite{ansari2}] that the total  flow of entropy, which is the difference of these two parts, corresponds exactly to physical quantities, more precisely to the full counting statistics of energy fluctuations.\cite{ansari2} This correspondence  is conceptually the analogue of the \emph{second law} for a quantum theory for thermodynamics, however limited to the weak coupling limits. Although the information content in the incoherent part  is carried out by standard correlations, the so-called Kubo-Martin-Schwinger (KMS) correlations\cite{KMS}, in the coherent part it indicates a large class of correlators that exists beyond the standard ones, the so-called \emph{extended} KMS correlations\cite{ansari1}.

\begin{figure}
\begin{center}
\includegraphics[scale=0.46]{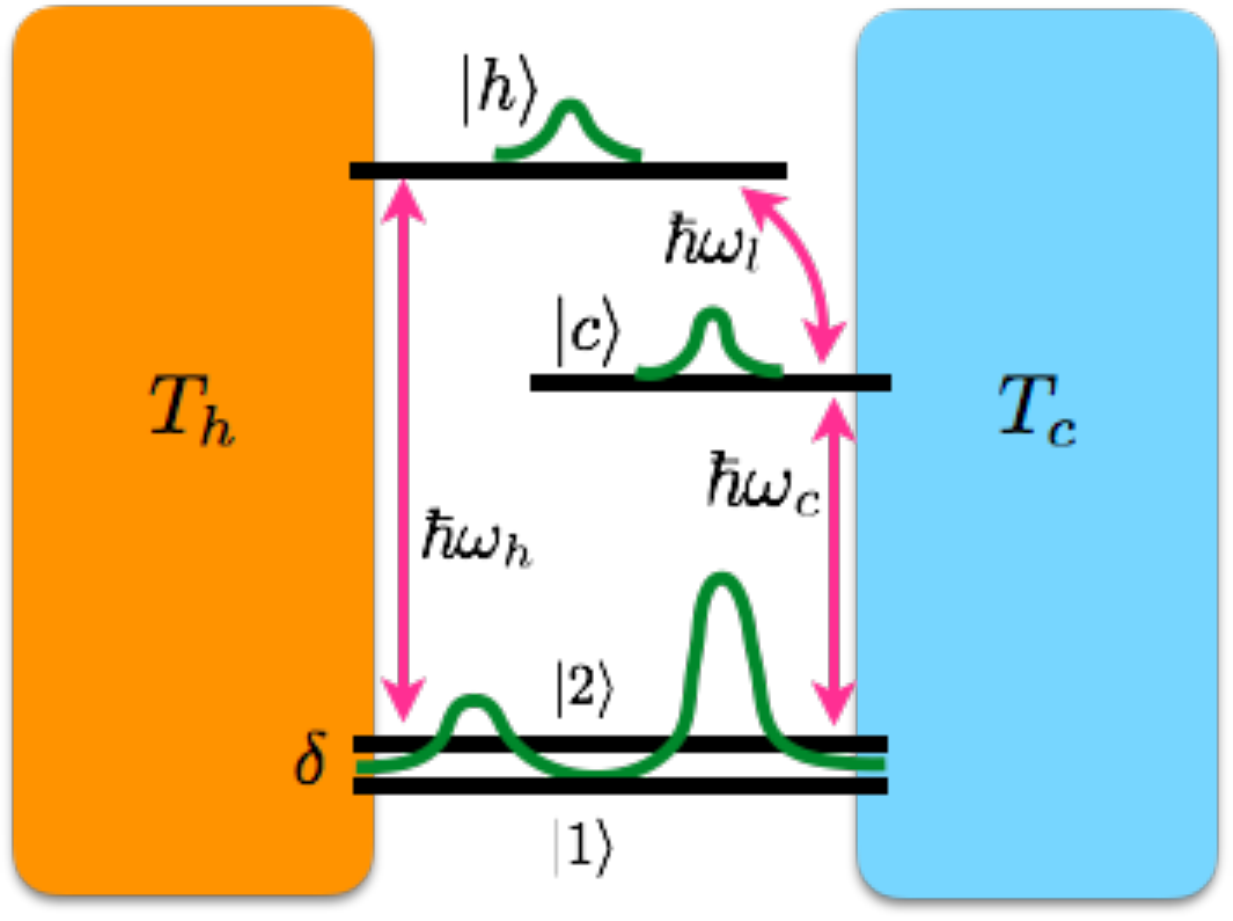}
\caption{(Color) A quantum heat engine with near degenerate ground states $|1\rangle$ and $|2\rangle$, and excited states $|h\rangle$ and $|c\rangle$, that resonantly (with frequency $\omega_l$) is coupled to two large heat baths kept at temperatures $T_c$ and $T_h$. Typical occupation probabilities are represented in green.}
\label{fig1}
\end{center}
\end{figure}

 In this paper we evaluate entropy flow for a practical QHE model compared to the simple modeled we previously studied. There are a number of QHE models that resemble interesting physical phenomena, such as light-harvesting biocells\cite{{cao},{whaley}},  photovoltaic cells\cite{{schullypc}, {khoshnegar},{kirk}}, lasing heat engines \cite{nikonov}. Our system of interest is the QHE introduced by Scully\cite{schullypc}, which has  4 energy level, two nearly degenerate ground state and two excited states, and is weakly coupled to two large heat baths kept at different temperatures $T_c$ and $T_h$, see Fig. (\ref{fig1}). This QHE in externally driven by a frequency that is almost equal to the energy difference of excited states.  The statistics of energy dissipations for this QHE has been previously studied in Ref.  [\onlinecite{Rahav}] and  shown to be non-Poissonian, from which various cumulants of energy exchanges can be extracted.  In this QHE several new phenomena have been studied, such as lasing without inversion \cite{schullylasing}, work extraction from single thermal reservoir \cite{schullysingle}, and elevated output powers \cite{schullypc}. Here, the main reason we  evaluate entropy flow  is to understand how to link energy fluctuations and entropy, and if the engine performs any sign of quantum thermodynamics beyond classical limits.  
   
  Our result shows interesting features in the entropy flow. We find that the presence of nonlinear flow can make entropy production in photovoltaic cells much slower or much faster than semiclassical rate.  We also study how decoherence time, which is induced by environments \cite{ansariqp}, and energy detuning, as a result of lifting degeneracy,  affect the net entropy flow. By designing a QHE with lower  decoherence time, we can speed up the flow of entropy between hot and cold baths. Lifting the degeneracy will result in the suppression of nonlinear flow that is the direct consequence of quantum coherence reduction.   Finally we obtain a lower bound on entropy flow as a direct result of nonlinearity in the flow.

In section (\ref{model}) we discuss the Hamiltonian  and the entropy evaluation to become prepared for section (\ref{qhe}) where we compute the flow in a 4-level QHE. Results are briefly discussed in section (\ref{discussion}) and some details can be found in Appendices A and B.

\section{The Model and Formalism}
\label{model}

In this section after introducing the Hamitonian model and the time evolution of density matrix, we explain how entropy as a quantity whose evaluation needs the time-evolution of a nonlinear operator can be consistently evaluated.

\subsubsection{The Hamiltonian}
Let us consider a QHE with quantum states $|x\rangle$ corresponding to energy eigenvalues $E_x$ coupled to a number of heat baths labeled by $\alpha$.  The Hamiltonian of this system is $H=H_{0}+H_{int}$ with non-interacting part $H_0$ being $H_{sys}+\sum_{\alpha} H_\alpha$ with the following system and reservoir Hamiltonians:  
\beqr \nonumber
\hat{H}_{sys}=\sum_{x} E_x|x\rangle \langle x|, && \\ 
\hat{H}_{\alpha}= \sum_{\bm{q}}\hbar \omega_{{\bm q},\alpha} \hat{b}_{{\bm q},\alpha}^\dagger \hat{b}_{{\bm q},\alpha}, &&
\label{eq.H0}
\eeqr
 and with $\hat{b}_{{\bm q},\alpha}$ ($\hat{b}_{{\bm q},\alpha}^\dagger$) being annihilation (creation) photon operator with  momentum ${\bm q}$ in the reservoir  $\alpha$. The interaction Hamiltonian is 
\beqr
\label{eq. V} \nonumber
&& \hat{H}_{int}= \sum_\alpha \sum_{xx'} \left|x\rangle\langle x'\right|X^{(\alpha)}_{xx'}\left(t\right) ,\\
&& \hat{X}^{(\alpha)}_{xx'}(t)=\hbar \sum_{{\bm{q}}}c_{xx',{\bm{q} \alpha }}\hat{b}_{{\bm{q} \alpha}}\exp(-i\omega_{{\bm{q} \alpha}} t)+h.c. 
\eeqr
with $\hat{X}$ operator acting on heat baths, complex-valued  $c_{xy,q\alpha}$ that couples the transition $x \to x'$ to a photon of certain momentum in a heat bath.  The coefficient $\exp(\pm i\omega_{{\bm{q} \alpha}} t)$ shows the time dependence of the creation and annihilation operators. We assume adiabatic switching on interaction such that far in the past $t\to -\infty$, all couplings are absent,  therefore the total density matrix is separable into subsystems. As the couplings slowly grow, density matrix can be formally determined from \cite{{nazarovbook}}
\beq
\hat{\rho}(t)=Te^{i\int_{\infty }^t d\tau \hat{H}_{int}(\tau)}\hat{\rho}(-\infty) \bar{T}e^{i\int_{-\infty}^t d\tau \hat{H}_{int}(\tau)},
\label{rhot}
\eeq
with $T$ ($\bar{T}$) being (anti-) time ordering operator.  One can expand Eq. (\ref{rhot})  in terms of $\hat{X}$ operators.  The Keldysh formalism\cite{Keldysh} is a general method to evaluate all energy orders\cite{{Keldysh},{Kamenev}}, for which the Keldysh contour is considered to represent the evolution of bra and ket states at different times; i.e. the ket (bra) states evolve along (opposite to) the time flow. Details can be found in Ref. [\onlinecite{nazarov11}].

\subsubsection{Entropy}

As mentioned above, entropy is nonlinear  in the density matrix of \emph{world}. Let us consider that in a world with the density matrix $\rho_w$ consisting of several systems and heat baths,  the system of  interest has the partial density matrix $\rho$.  
The entropy of this system is  $S=-\textup{Tr}\hat{\rho}\ln \hat{\rho}$ and evaluated by tracing out all except the system of interest. Here we assume $k_B=1$. This  logarithmic dependence makes the evaluation of entropy mathematically involved. 

Consider the simple example that  interactions are so weak that the quantum system is only perturbed in the vicinity of its equilibrium state at  ground state. In this case the density matrix can be approximated to  $\rho(t)\approx p_0 + \rho^{(1)}(t)$ with $|\rho^{(1)}/p_0|\ll 1$. The entropy flow, i.e. ${F}=dS/dt$, becomes  $ -(1+ \ln p_0) d\rho^{(1)} /dt - (1/2p_0) d(\rho^{(1)})^2/dt+\cdots $, which is clearly nonlinear in density matrix. We showed in Ref. [\onlinecite{ansari1}] that for any positive $n$  one can show $d(\rho)^n/dt \neq  n(\rho^{n-1}) d\rho/dt$. In other words, one cannot simplify $- (1/2p_0) d(\rho^{(1)})^2/dt$ to  $- p_0 d\rho^{(1)}/dt$. Such a simplification is only meaningful for noninteracting  systems.

First one must notice that  in the logarithmic expansion of entropy there are infinite terms to be computed. This is impossible and moreover we cannot find any clear criteria to make a meaningful truncation on the expansion. Nazarov in Ref. [\onlinecite{nazarov11}] suggested that we rewrite entropy as a limit of the Renyi entropies\cite{{renyi}}, i.e. $S=-\lim_{M\to 1} dS_M/dM$ with the Renyi entropy of positive degree $M$ being $S_M=\textup{Tr}_r \{\rho_r\}^M$. Naturally computing the Renyi entropy flows is the next problem to achieve, see Appendix (\ref{appA}).  We proposed in Ref. [\onlinecite{ansari1}] how to compute the time evolution of the operator $\{\hat{\rho_r}\}^M(t)$ without using the solution of $\hat{\rho}$. In order to evaluate the entropy flow in a quantum system one should
evaluate the Renyi entropy flow  and analytically continue it to $M\to 1$. Detailed analysis shows that the consistent entropy flow has two parts $Q_i$ and $Q_c$:
\beqr \nonumber 
&& 
\frac{dS}{dt} =\frac{ Q_i-Q_c}{T}\\ 
\nonumber
&& Q_{i}   =   \sum_{x'yy'}\rho_{x'y}  \tilde{\chi}_{y'x',yy'}\left(\omega_{yy'}\right)   \bigg(\bar{n}(\omega_{yy'})+1\bigg) \omega_{yy'}  \\ 
&& Q_{c}   =    \sum_{xx'yy'}\rho_{x'x}\rho_{y'y}  \tilde{\chi}_{xx',yy'}\left(\omega_{yy'}\right)   \omega_{yy'}
\label{eq. entfw}
\eeqr
with $Q_i$ being the incoherent flow of entropy and $Q_c$ the coherent part. $\hbar \omega_{yy'}\equiv E_y-E_{y'}$ and $\tilde{\chi}_{xx',yy'}$ is the generalized dynamical susceptibility between two transitions:  $|x\rangle \to |x'\rangle$ and $|y\rangle \to |y'\rangle$.  $\bar{n}$ denotes the Bose distribution.

 Eq. (\ref{eq. entfw}) has two parts: 1) the incoherent flow  $Q_i$, which  is linearly proportional to reduced density matrix, and 2) the coherent part $Q_c$, which is nonlinear--in fact quadratic because we calculate it in the second order perturbation theory. This entropy flow can be equivalently determined from using the corresponding physical quantities, which are the full counting statistics of energy transfers, see Appendix (\ref{appB}).

\section{Four-level Quantum Heat Engines}
\label{qhe}

Let us calculate entropy flow in the four-level quantum heat engine (QHE) introduced by Scully \emph{et al.}\cite{schullypc} and shown in Fig. (\ref{fig1}). This QHE consists of two nearly degenerate lower levels $|1\rangle$  and $|2\rangle$ (denoted by label $i, j=1,2$) with energy $E_1= E_2+\delta$ and $\delta$ being energy detuning, and two excited levels $|h\rangle$ and $|c\rangle$ with energies $E_h$ and $E_c$ (denoted by labels $\alpha=h,c $), see Fig. (\ref{fig1}). An example of such QHE is a laser heat engines in which environmental noise  helps to increase net emitted laser.  The full counting statistics of energy transfers in this QHE has been calculated in Ref. [\onlinecite{Rahav}]. By applying the first cumulant of the energy statistics in the second law (as we will show in next section) we can immediately determine a semiclassical value for the entropy flow. However, here we calculate it using a formalism that is free of any assumption about the underlying quantum thermodynamics. Therefore  we notice that our results are dramatically different from what semiclassical approach predicts.

Here we choose the probe environment to be the hot bath with temperature $T_h$. The quantum system is externally driven by a single-mode cavity at the frequency $\omega_l\approx (E_h-E_c)/\hbar$. The driving  Hamiltonian is $\hat{H}_{sys-dr}=\Omega (\hat{b}_l^\dagger |c\rangle\langle h|+\hat{b}_l |h\rangle\langle c|)$,  $\hat{b}_l$ ($\hat{b}_l^\dagger$) being the annihilation (creation) operator for the cavity mode; $\langle \hat{b}^\dagger_l \hat{b}_l\rangle = \bar{n}_l $;  $\langle \hat{b}_l \hat{b}^\dagger_l \rangle = \tilde{n}_l$ with $\bar{n}_l$ being the average number of photons in the cavity and $\tilde{n}_l\equiv \bar{n}_l+1$.  power.\cite{{schullypc},{schullylasing}}  

The stationary density matrix solution can be summarized in the vector $\bm{R}=\{\rho_{11}, \rho_{22}, \rho_{hh}, \rho_{cc}, \textup{Re}(\rho_{12}) \}$. This vector can be evaluated using the time evolution equation $d\bm{R}/dt=\mathscr{L}\bm{R}$ equation, by assuming that the density matrix slowly varies (Markov approximation), thus we approximate $\rho(t-t')\approx \rho(t)$. One can determine the following  $\mathscr{L}$ for the dynamics: 
\begin{eqnarray} \nonumber
\label{eq.L}
&& \mathscr{L}\equiv \\ && \left(\begin{array}{ccccc}
\chi_{11} & 0 & \tilde{\chi}_{h1}\tilde{n}_{h} & \tilde{\chi}_{c1}\tilde{n}_{c} & -2\tilde{\chi}_{12}\\
0 & \chi_{22} & \tilde{\chi}_{h2}\tilde{n}_{h} & \tilde{\chi}_{c2}\tilde{n}_{c} & -2\tilde{\chi}_{12}\\
\tilde{\chi}_{h1}\bar{n}_{h} & \tilde{\chi}_{h2}\bar{n}_{h} & \chi_{hh} & \Omega^{2}n_{l} & 2\tilde{\chi}_{1h,h2}\bar{n}_{h}\\
\tilde{\chi}_{c1}\bar{n}_{c} & \tilde{\chi}_{c2}\bar{n}_{c} & \Omega^{2} \tilde{n}_{l} & \chi_{cc} & 2\tilde{\chi}_{1c,c2}\bar{n}_{c}\\
-\tilde{\chi}_{12} & -\tilde{\chi}_{12} & \tilde{\chi}_{1h,h2}\tilde{n}_{h} & \tilde{\chi}_{1c,c2}\tilde{n}_{c} & \chi
\end{array}\right)\nonumber \\ &&
\end{eqnarray}
with  coupling $\chi_{ii}$ being $ -\sum_{\alpha} \tilde{\chi}_{i\alpha,\alpha i} \bar{n}_i$;   $\chi_{hh}$ being $ -\sum_{i} \tilde{\chi}_{ih,h i} \tilde{n}_h-\tilde{n}_l\Omega^2$;   $\chi_{cc}$ being $ -\sum_{i} \tilde{\chi}_{ic,c i} \tilde{n}_c- \bar{n}_l \Omega^2$;  $\chi$ being $-(1/2)\sum_{i,\alpha}\tilde{\chi}_{i\alpha,\alpha i} \bar{n}_\alpha -1/\tau_2$;  and the symmetric $\chi_{ij}$ being $ (1/2)\sum_{\alpha}\tilde{\chi}_{i\alpha,\alpha j} \bar{n}_\alpha$ for $i\neq j$.  

This evolution equation has been solved explicitly, see Eq. (S26) of the Supplementary Materials in Ref. [\onlinecite{schullypc}].  One should notice that $\chi$ depends on decoherence time $\tau_2$, which is induced by environment and affects all elements of the system density matrix, including an exponential decay of the imaginary part of the coherence $\textup{Im} \rho_{12}\sim \exp(-t/\tau_2)$. 

Being equipped with the  stationary solution for $\bm{R}$, we can now compute the entropy flow by substituting it in Eq. (\ref{eq. entfw}). The  result is that the stationary flow of entropy from the probe (hot) heat bath becomes: 
\beqr \nonumber
  \frac{dS}{dt} & =&  \bigg\{      {\gamma}    p_h -  E_{h2} \tilde{\chi}_{h2} \bar{n}\left(\frac{E_{h2}}{T_h}\right) p_2   -  E_{h1} \tilde{\chi}_{h1} \bar{n}\left(\frac{E_{h1}}{T_h}\right)  p_1  \bigg.  
\\  && \nonumber  \ \  
 -   \tilde{\chi}_{1h,h2} \left[ E_{h1} \bar{n}\left(\frac{E_{h1}}{T_h}\right) + E_{h2} \bar{n}\left(\frac{E_{h2}}{T_h}\right)\right] \textup{Re}\rho_{12}      \nonumber
 \\  &&  \bigg. - \frac{1}{2} \sum_{i=1,2}  E_{hi} \tilde{\chi}_{1h,h2} |\rho_{12}|^2     \bigg\} / T_h 
 \label{eq. rflow ex}
\eeqr
with $p_x\equiv \rho_{xx}$ for $x$ being $1,2,h,c$.  $\tilde{\chi}_{\alpha i}\equiv \tilde{\chi}_{i\alpha,  \alpha i} (\omega_{i\alpha})$ being the dynamical response function, and   $\tilde{\chi}_{1\alpha,\alpha 2}=\sqrt{\tilde{\chi}_{\alpha 1} \tilde{\chi}_{\alpha 2}}$. Moreover ${\gamma}\equiv \sum_{i=1,2 } \left[\bar{n}\left({E_{hi}}/{T_h}\right)+1\right] E_{hi} \tilde{\chi}_{hi} $.  Equivalently one can compute the flow of entropy by using its corresponding  full counting statistics of energy transfers in Eq. (\ref{eq. rfcs}).

 In Eq. (\ref{eq. rflow ex}) the linear terms determine incoherent flow, where the first term is the entropy gain by photon  absorption and the next three terms are entropy loss in photon decays. The quadratic terms  is the coherent flow that is the entropy loss due to  extended KMS correlators.  
 
 Before discussing  results, let us briefly show how one can derive the incoherent part independently using the combination of textbook second law  and the statistics of  energy transfers worked out in  Ref. [\onlinecite{Rahav}]. The full counting statistics generating function of energy  $E_{tr}$ being exchanged during time interval $\mathcal{T} $ is  $G(\xi, t) = \textup{Tr} \rho(\xi,t)$ with $\rho(\xi,t)$ being the stationary solution of $d\rho(\xi,t)/dt=\mathscr{L}(\xi) \rho(\xi,t)$, with  $\xi$ being the characteristic parameter. The explicit form of $\mathscr{L}(\xi)$ is given by Eq. (14) of Ref. [\onlinecite{Rahav}]. This generating function determines the first cumulant $d\langle E \rangle /dt$. By substituting it in the relation between entropy and energy flows  $dS/dt=(1/T) dE/dt$, one can exactly obtain the first two lines of  Eq. (\ref{eq. rflow ex}). However this is important to notice that the second law is unable to provide the complete flow of entropy with coherent flow included.  
 
\begin{figure}
\begin{center}
\includegraphics[scale=0.45]{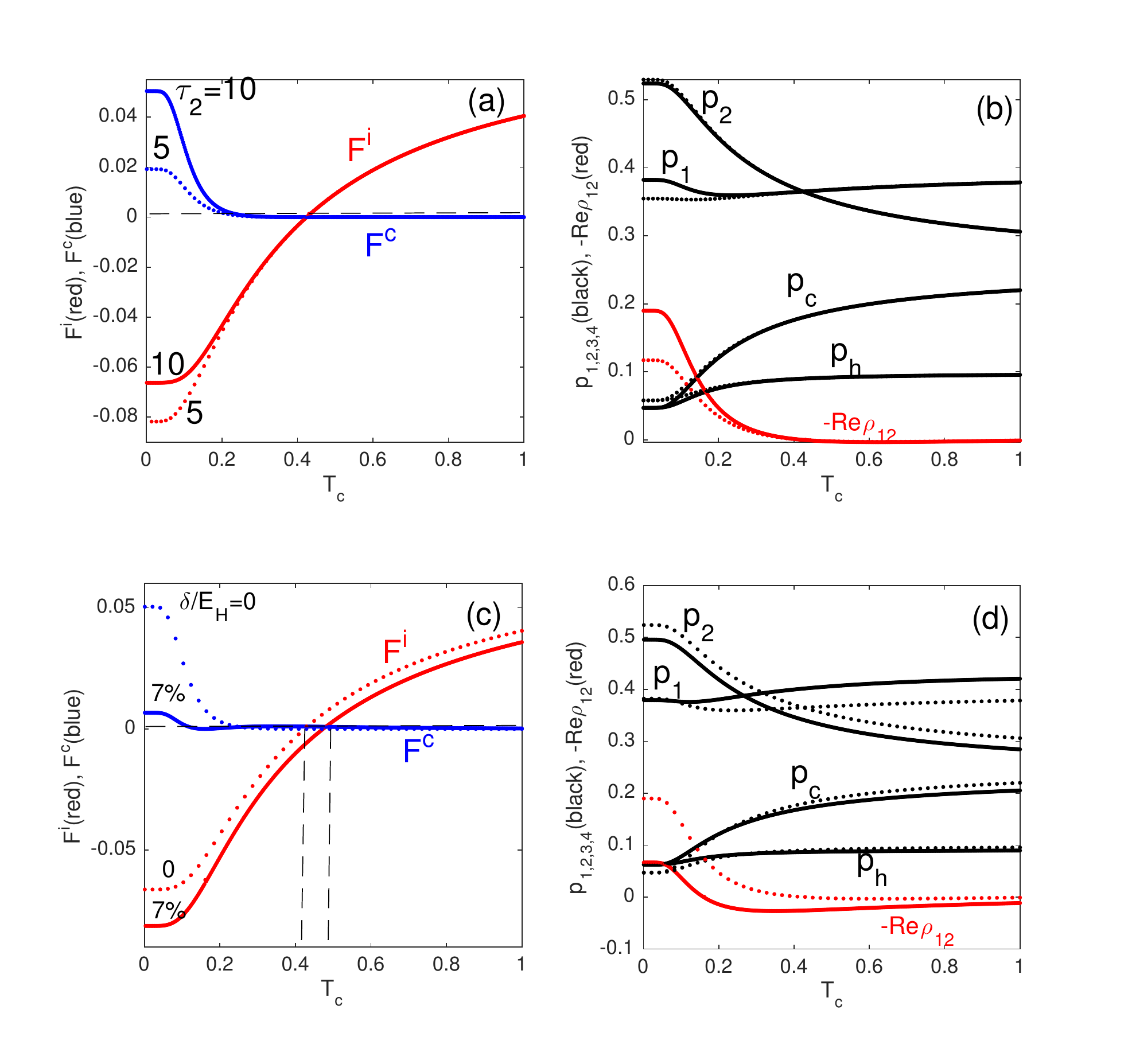}
\caption{(Color)  Incoherent (red) and coherent (blue) entropy flows in hot (probe) environment versus cold bath temperature $T_c$ for (a) two different values of decoherence time $\tau_2=$5 (dotted) and 10 (solid) and the absence of detuning energy, (c) two different values of detuning $\delta/E_{h1}=$0 (dotted) and 7\% (solid) both for the case of longer detuning time $\tau_2=$10. (b,d) show the corresponding  $p_1, p_2, p_c, p_h$ (black) and $-\textup{Re}\rho_{12}$ (red) with the same order for dotted and solid lines.   Other parameters are   $E_h=1.5$, $E_c=0.4$, $E_1=E_2=0.1$, $T_h=\Omega/2=n_l=\tilde{\chi}_{h1}=\tilde{\chi}_{h2}=\tilde{\chi}_{c2}=10 \tilde{\chi}_{c1}=1$. }
\label{fig2}
\end{center}
\end{figure}

Let us consider two QHEs with no energy detuning, i.e. $\delta=0$, and  for two different values of decoherence time, i.e. $\tau_2=5$ (dotted) and 10 (solid). The incoherent (coherent) entropy flows $F^i$ ($F^c$) in probe environment depends on the temperature of the cold environment $T_c$.  This dependence is plotted in Fig. (\ref{fig2}a-d) for the couplings specified in its caption. The net entropy is ${F}={F}^i-{F}^c$ as shown in Eq. (\ref{eq. entfw}).

At low $T_c \ll T_h$ Fig. (\ref{fig2}a) indicates that entropy flows out of the hot bath.   By warming up  cold bath above  the \emph{onset} temperature $T_o>0.42$ the overall ground state populations as shown in panel (b) are suppressed and the excited states become more populated. This causes the \emph{reversal} flow of entropy from the cold to hot bath. At the onset temperature  there is no flow of entropy expected. As shown the onset temperature does not depend on decoherence time. Let us now compare the net entropy flow $F=F^i-F^c$ determined from the consistent formalism and the semiclassical flow predicted by  the second law, which is equivalent to the incoherent part $F^i$. One can see in Fig. (\ref{fig2}a) that the net flow is heavily suppressed. This suppression takes place due to the contribution of nonlinear (coherent) part of the flow that is quadratic in $\rho_{12}$ and is as important as the semiclassical part where  $\rho_{12}$ is not negligible.  Moreover at low  $T_c$ limit the smaller decoherence time $\tau_2$ is, the faster entropy flows out of the probe environment.

Now let us study the effect of energy detuning on  entropy flow. Fig. (\ref{fig2}c) shows incoherent (coherent) entropy flow $F^i$ ($F^c$) for two QHEs with detuning ratio $\delta/E_{h2}=$ 0 (dotted) and  7\% (solid). Panel (d) shows the corresponding stationary populations and coherence. By increasing energy detuning the onset temperature becomes larger. This is mostly because of a sign change in $\textup{Re}\rho_{12}$ in the presence of detuning, which in Eq. (\ref{eq. rflow ex}) makes the linear term on $\textup{Re}\rho_{12}$  turn from positive at low $T_c$ to negative at higher $T_c$ and this reduces the total entropy flow and causes a shift of the onset temperature forward.

Now let us simplify the entropy flow for the cold bath being at zero temperature limit.  One can see in Fig. (\ref{fig2}) that at zero $T_c$ both parts of entropy flow are at their extrema and with opposite sings. We use Eq. (\ref{eq. rflow ex}) and analytically compute the flow for an engine with zero detuning. One can  simplify the time evolution of $p_1$ and $p_2$  in the limit of $T_c \approx 0$ using Eq. (\ref{eq.L}).  Some lines of algebraic calculations shows that the stationary value of the ground state occupation probabilities under such conditions are $p_i=p_h \exp(E_H/k_BT_h) + (\eta_i/\bar{n}(E_H/T_h)) p_c - \lambda_i \textup{Re}{\rho}_{12}$ with $\eta_i$ being $\tilde{\chi}_{ci}/\tilde{\chi}_{hi}$;   $\lambda_1=\sqrt{r_h}$;  $\lambda_2=1/\lambda_1$;  $E_H\equiv E_{hi}$; and  $r_h\equiv \tilde{\chi}_{h2}/\tilde{\chi}_{h1}$.   Substituting them all in the entropy flow of Eq. (\ref{eq. rflow ex}) will determine the entropy flow for the QHE: 

\beq
\label{eq. flow0T}
\left.\frac{dS}{dt}\right|_{T_c\approx 0} = - \frac{E_H }{k_BT_h}\left( {   {p_c} (\tilde{\chi}_{c1}+\tilde{\chi}_{c2}) + \tilde{\chi}_{1h,h2} |\rho_{12}|^2 }\right)  
\eeq

 Eq. (\ref{eq. flow0T}) clearly states that the engine at zero cold bath temperature exhibits a persistent negative  flow of entropy from hot to cold bath, no matter what are other parameters. This, at least in the weak coupling limit, indicates no violation takes place against the third law in this engine. 
 
Given that the entropy flow changes sign at the onset temperature, one can simplify  Eq. (\ref{eq. rflow ex})  to find out the condition where reversal flow occurs, which is : 
\beqr  \nonumber
 \left[ \bar{n} \left({{E_H}/{T_h}}\right) +1\right] (1+{r_h}) p_h && \\ \nonumber   -  \bar{n} \left({{E_H}/{T_h}}\right) \left(p_1+ r_h p_2 + 2 \sqrt{r_h} \textup{Re}\rho_{12}  \right)  - \sqrt{r_h} |\rho_{12}|^2   & \geq 0& \\ 
 \label{eq.neg}
\eeqr
with zero flow at the  onset temperature.

Before concluding, let us make some important remarks about the net entropy flow in these quantum photovoltaic cells. For devices with $r_h=1$,  by dividing both sides of Eq. (\ref{eq. rflow ex}) by $\tilde{\chi}_{h1}E_H$ and denoting  left side  $f=(dS/dt)/\tilde{\chi}_{h1}E_H$, a few lines of algebra simplifies the result into a quadratic equation for quantum coherence:  $|\textup{Re}\rho_{12}|^2+2\bar{n} \textup{Re}\rho_{12} + \bar{n}(1-p_c) - (3\bar{n}+2) p_h  + f = 0$. Solving this equation for $\textup{Re}\rho_{12}$ will determine the condition for  it to be  real-valued.  One can  find the  forbidden zone  is  where $p_c+(3\bar{n}+2)/\bar{n} p_h < f-\bar{n} + 1$. Given that the left side of this inequality is positive-valued, the left side cannot be negative and therefore the following lower bound on net entropy flow holds: 
$dS/dt \geq (\bar{n}(E_H/T_h)-1)E_H \tilde{\chi}_{h1}$. In the limit of $E_H/T_h \ll 1$ this entropy lower bound can be approximated by $dS/dt \geq  1- 3E_H/2T_h + O(2)$.     Notice that the existence of a lower bound  relies on the  quadratic dependence of entropy flow on the quantum coherence. Such a quadratic dependence is absent in semiclassical analysis and therefore no lower bound is expected.  

The entropy flow we obtained  here for this QHE can be measured experimentally using its exact corresponding partners in physical quantities. These physical quantities are   the full counting statistics of energy transfers. This interestingly indicates that the entropy lower bound reveals the existence of  a corresponding constraint on energy fluctuations in the system. This can be further developed and  experimentally tested.

\section{Summary}
\label{discussion}

We calculated the entropy flow of a 4-level quantum heat engine within weak interaction limit. The results,  obtained from the full quantum formalism of multiple parallel worlds,  show that in addition to  semiclassical results, entropy flow has nonlinear contribution of quantum coherence as the result of coherent drive. The presence of this nonlinear term significantly suppresses the semiclassical value for entropy, however this heavy suppression does not allow for entropy to flow out of a cold bath at zero temperature; thus the third law is not violated in the weak coupling regime.  We also explicitly determined the by reducing  environmentally-induced decoherence time the onset temperature does not change but the flow can take place much faster.  Lifting the degeneracy will result in the suppression of quantum coherence that directly reduces the nonlinear term of entropy.  Finally we argued that the quadratic dependence of entropy flow on quantum decoherence, which is absent in semiclassical analysis,  determines a lower bound on entropy flow. Given that there is an exact correspondence between entropy flow and energy fluctuations, one can expect that the lower bound on entropy flow can correspond to a constraint on energy fluctuations, which can be the subject of future research.

Let us now discuss how to measure the entropy. The entropy flow is not accessible in direct measurement as they are nonlinear functions of density matrix.   Direct measurements of density matrix for a probe environment requires characterization of reduced density matrix of an infinite system, which is a rather nontrivial procedure and needs the complete and precise re-initialization of the initial density matrix. Measuring  entropy flow from their physical  correspondence requires that some generating functions are extracted from determining statistical cumulants of transferred energy in experimental data.  The measurement procedures may be complex, yet doable and physical. 

Our derivation was restricted to the second order perturbative dynamics. Let us briefly explain how this physics can be extended to strong coupling regime.  Below I will describe two approaches for the development:  One can use the polaron transformation\cite{legett} to incorporate the high-order system-bath interaction into the system dynamics. This transformation will change the generalized correlators of  heat baths as well as the Renyi entropies.  Alternatively, one can define the generalize density matrix $\mathcal{R}$ to include the density matrix of $M$ worlds, and extend the dynamical equation for  $\mathcal{R}(t)$. The solution is a set of eigensolutions proportional to $\mathcal{R}(t) \approx \exp (-\Gamma t )$. In strong coupling limit there is no stationary solution with zero $\Gamma$, instead the flow of Renyi entropy is $\mathscr{F}_M=\Gamma_{0}$ with $\Gamma_{0}$ being the  closest eigenvalue to zero. This will help to identify the entropy flow in the limit of $M\to 1$ and is the subject of ongoing research.

\appendix

\section{Renyi entropy flow}
\label{appA}
Evaluating R\'{e}nyi entropies requires time evolution of integer powers of density matrix.   Consider a closed system with total density matrix $\rho$ made of two interacting systems $A$ and $B$. The reduced density matrix for system $A$ is  $\rho_A=\textup{Tr}_{B} \rho$. The Renyi entropy of degree $M$ in the system $A$ is  $\ln S_M^{A}=\ln \textup{Tr}_A \{( \rho_A)^M\}$. If the two systems do not interact, the entropies are conserved $d\ln S_M^{A,B}/dt=0$; however for interacting heat baths in thermal equilibria, a steady flow of entropy is expected from one heat bath to another one. This is similar to the steady flow of charge in an electronic junction that connects two leads kept at different chemical potentials \cite{ansariqp}. Defining the Renyi entropy flow of degree $M$ in system $A$ as $\mathscr{F}_M^{A}=d\ln S_M^{A}/dt$,  there is a conservation law for  $\mathscr{F}_M^{A+B}$; however due to the inherent non-linearity $\mathscr{F}_M^{A}+\mathscr{F}_M^{B}\neq 0$ and equality holds only approximately, subject to volume dependent terms.\cite{nazarov11}

For the evaluation of Renyi entropy flow of degree $M$ in the system $A$, i.e. $d\ln S^{A}_M/dt$, in the second order perturbation we need to compute $d (\rho_A)^M/dt$. Let us drop the index $A$ from the equation from now. Considering the initial density matrix is $\rho_0$ and it can be found later to take the following value $\rho(t) = \rho_0 +\rho^{(1)}(t)+O(2)$, with  $\rho^{(1)}=\rho^{(1)}_I+\rho^{(1)}_{II}$ and $\rho^{(1)}_{I}(t) =- i \int_0^t dt' \hat{H}(t')\hat{\rho}(t)$ and $\rho^{(2)}_{II}(t) = i \int_0^t dt' \hat{\rho}(t) \hat{H}(t')$. The superscripts of parenthesis indicates perturbation order.  Similarly $d\rho/dt = \dot{\rho}^{(1)}+\dot{\rho}^{(2)} + O(3)$ with $\dot{\rho}^{(1)}  =\dot{\rho}^{(1)}_{1} + \dot{\rho}^{(1)}_{2}$ and  $\dot{\rho}^{(1)}_{1}(t) =- i  \hat{H}(t)\hat{\rho}$ and $\dot{\rho}^{(1)}_{2}(t) = i \hat{\rho} \hat{H}(t)$ and $\dot{\rho}^{(2)} = i[\dot{\rho}^{(1)},H]$. The flow of nonlinear measure can be expanded as follows: $d(\rho)^M/dt=(d\rho/dt)(\rho)^{M-1}+\rho (d\rho/dt)(\rho)^{M-2}+\cdots+(\rho)^{M-1}(d\rho/dt)$. Using these definitions we can evaluate  $d\rho^M/dt$ in the second order as follows:
\beqr
\label{eq. 1} \nonumber
\frac{d\rho^M}{dt} &=& \left\{  \dot{\rho}^{(2)} \rho_0^{M-1} +  \rho_0 \dot{\rho}^{(2)}  \rho_0^{M-2}+ \cdots+\rho_0^{M-1} \dot{\rho}^{(2)} \right\} + \nonumber \\ && \nonumber 
\bigg\{ \dot{\rho}^{(1)} \left[ \rho^{(1)} \rho_0^{M-2} + \rho_0 \rho^{(1)} \rho_0^{M-3} + \cdots\right] \bigg. \\ \nonumber
&& \ \ \  + \rho_0 \dot{\rho}^{(1)} \left[ \rho^{(1)} \rho_0^{M-3}  \rho_0 \rho^{(1)} \rho_0^{M-4}+\cdots\right]  \\  && \ \ \    +    \cdots+ \bigg. \rho_0^{M-2} \dot{\rho}^{(1)}   \rho^{(1)} \bigg\}  \eeqr
where the first line of Eq. (\ref{eq. 1}) represent  photon exchanges taking place only within one world, and the remaining terms represent  the exchange  of photons between different copies of world density matrices.

We implement the extended Keldysh formalism for the analysis of Renyi entropy flow. Detailed analysis with all diagrams that can be seen in  Appendix B of [\onlinecite{ansari1}]. Rigorous analysis shows that the Renyi entropy flow is:

will result the following flow of Renyi entropy:
\beqr \nonumber 
&& \mathscr{F}_M   = \sum_{yy'} \frac{M \bar{n} \left(M\omega_{yy'}\right)}{\bar{n}\big(\left(M-1\right)\omega_{yy'}\big) \bar{n}\left(\omega_{yy'}\right)  \omega_{yy'}}  \left\{ Q_{yy'}^{i} - Q^{c}_{yy'}\right\} \\ 
\nonumber
&& Q^{i}_{yy'}   =   \sum_{x'}\rho_{x'y}  \tilde{\chi}_{y'x',yy'}\left(\omega_{yy'}\right)   \bigg(\bar{n}(\omega_{yy'})+1\bigg) \omega_{yy'}  \\ 
&& Q^{c}_{yy'}   =    \sum_{xx'}\rho_{x'x}\rho_{y'y}  \tilde{\chi}_{xx',yy'}\left(\omega_{yy'}\right)   \omega_{yy'}
\label{eq. rflow}
\eeqr
with $\hbar \omega_{yy'}\equiv E_y-E_{y'}$.

In Eq. (\ref{eq. rflow}) there are two types of flows contributing: (i) the incoherent flow  $Q^{i}$, for quantum leaps on energy levels, and (ii) the coherent flow $Q^{c}$ for the exchange of energy through the quantum coherence. $Q^i$ is in fact represent the first line of Eq. (\ref{eq. 1}) and $Q^c$ to the rest of them.

\section{Full counting statistics of energy dissipations}
\label{appB}

This correspondence makes evaluation of the Renyi entropy flows possible using full counting statistics of energy exchanges.\cite{ansari3} Previosuly similar correspondence has been found in charge transport in Ref. [\onlinecite{{klish},{flindt}}] 

Let us briefly recall what is the Full Counting Statistics (FCS) of energy transfers between a small quantum system weakly coupled to a classical environment kept at temperature $T$.\cite{Kindermann} It concentrates on the probability $\mathscr{P}^{(T)}(E_{tr} ,\mathcal{T} )$ to have energy $E_{tr}$ transferred between two systems during time interval $\mathcal{T} $.\cite{Kindermann} The superscript $(T)$ refers to the temperature of environment.  In the long $\mathcal{T} $ limit all statistical cumulants of the
energy transfers can be determined from  
the generating function $G^{(T)}(\xi) = \int_{0}^T dE_{tr} \mathscr{P}^{(T)}(E_{tr},\mathcal{T}  ) \exp(i \xi E_{tr}) \approx \exp[- \mathcal{T}  f^{(T)}(\xi)]$. The parameter $\xi$ is a characteristic parameter and cumulants are given by expansion of $f(\xi)$ in $\xi$ at $\xi=0$.

The correspondence between Renyi entropy and FCS of energy transfers can be further simplified to evaluate directly the flow of von Neumann entropy between a small quantum system weakly coupled to a classical environment kept at temperature $T$ using the following formula:
\beq
\label{eq. rfcs}
\frac{dS}{dt} = 
\lim_{M\to 1} M\left\{f^{\left(\frac{T}{M}\right)}\left(\frac{1-M}{iT}\right) - \overline{f}^{\left(\frac{T}{M}\right)}\left(\frac{1-M}{iT} \right)\right\}
\eeq

In the right side of Eq. (\ref{eq. rfcs}) there are two generating functions that should be evaluated at rescaled temperature $T/M$ and non-zero parameter $\xi=(1-M)/iT$;  $f$ is the generating function by means of interaction between quantum system and  environment, where $\bar{f}$ is an auxiliary generating function statistics that carries only the coherent exchange of energy between environment and quantum system.  To understand $\bar{f}$ let us consider that the interaction Hamiltonian is $\hat{H} = \hat{X}\hat{Y}$ with $\hat{X}$ acting on classical environment and $\hat{Y}$ on quantum system.  This FCS generating function is associated to a Hamiltonian in which averaging takes place over the system part; i.e. $\hat{Y} \to  \langle \hat{Y}\rangle$. This will be the Hamiltonian of the equilibrium system subject to time-dependent external forces. In Ref. [{\onlinecite{ansari2}]  we discussed the physical realization of the scheme. We showed that Eq. (\ref{eq. rfcs}) can provides the textbook second law of thermodynamics in the absence of quantum coherence.


\begin{thebibliography}{99}
\bibitem{Horodecki} R. Horodecki, M. Horodecki, K. Horodecki, Rev. Mod. Phys. 81, 865 (2009).
\bibitem{vidal} G. Vidal, J. I. Latorre, E. Rico, and A. Kitaev, Phys. Rev. Lett. 90, 227902 (2003)
\bibitem{dur} W. Dur, L. Hartmann, M. Hein, M. Lewenstein, H.J. Briegel, Phys. Rev. Lett. 94, 097203 (2005).
\bibitem{clarke} J. Clarke and F. Wilhelm, Nature 453, 1031 (2008)
\bibitem{wendin}  G. Wendin and V. Shumeiko, in Handbook of Theoretical and Computational Nanotechnology, edited by M. Rieth and W. Schommers (American Scientific Publishers, Valencia, CA,(2006), Chap. Superconducting Quantum Circuits, Qubits and Computing.
\bibitem{wilhelm} J.M. Martinis, M. Ansmann, J. Aumentado, Phys. Rev. Lett. 103, 097002 (2009); M.H. Ansari, FK Wilhelm, U Sinha, A Sinha, Superconductor Science and Technology 26 (12), 125013 (2013), arXiv:1211.4745 
\bibitem{bal} S. Gustavsson,  F. Yan, G. Catelani, J. Bylander, A. Kamal, J. Birenbaum, D. Hover, D. Rosenberg, G. Samach, A.P. Sears, S.J. Weber, J.L. Yoder, J. Clarke, A.J. Kerman, F. Yoshihara, Y. Nakamura, T.P. Orlando, W.D. Oliver, Science 354, 1573–1577 (2016); M. Bal, M.H. Ansari, J.L. Orgiazzi, R.M. Lutchyn, A. Lupascu,  Physical Review B 91 (19), 195434 (2015)  arXiv:1406.7350.
\bibitem{seifert} C. Tietz, S. Schuler, T. Speck, U. Seifert, J. Wrachtrup, Phys. Rev. Lett. 97, 050602 (2006); U. Seifert, Rep. Prog. Phys. 75, 126001 (2012); U. Seifert, Phys. Rev. Lett. 95, 040602 (2005).
\bibitem{toyabe} S. Toyabe, T. Sagawa, M. Ueda, E. Muneyuki, M. Sano, Nat. Phys. 6, 988 (2010). 
\bibitem{esposito} M. Esposito, Phys. Rev. E 85, 041125 (2012).
\bibitem{koski} J. V. Koski, T. Sagawa,	O-P. Saira,	Y. Yoon, A. Kutvonen, P. Solinas, M. Möttönen, T. Ala-Nissila, J. P. Pekola, Nature Physics 9, 644-648  (2013).
\bibitem{pekola} J.P. Pekola, Nature Physics 11, 118-123 (2015).
\bibitem{schullypc} M.O. Scully, K.R. Chapin, K.E. Dorfman, M.B. Kim,  A. Svidzinsky, Proc. Natl. Acad. Sci. USA, 108, 15097 (2011).
\bibitem{Maruyama} K. Maruyama, F. Nori, V. Verdal, Rev. Mod. Phys. 81, 1 (2009).
\bibitem{Dorfmana} K.E. Dorfmana, D.V. Voroninea, S. Mukamelc, M.O. Scully, Proc. Natl. Acad. Sci. USA 2746 (2013).
\bibitem{sambur} J.B. Sambur, T. Novet, B.A. Parkinson, Science 330, 63 (2010).
\bibitem{cho} B. Cho, W.K. Peters, R.J. Hill, T.L. Courtney, D.M. Jonas, Nano Lett. 10, 2498 (2010).
\bibitem{schaller} R.D. Schaller, V.I. Klimov, Phys. Rev. Lett. 92, 186601 (2004); R. Baer and E. Rabani, Chem. Phys. Lett. 496, 227 (2010).
\bibitem{jar} C. Jarzynski, Phys. Rev. Lett. 78, 2690 (1997).
\bibitem{crooks} G.E. Crooks, Phys. Rev. E 60, 2721 (1999).
\bibitem{Amico} L. Amico, R. Fazio, A. Osterloh, and V. Vedral, Rev. Mod. Phys. 80, 517 (2008). 
\bibitem{wilde} M. Berta, K. P. Seshadreesan, M. M. Wilde,  Phys. Rev. A 91, 022333 (2015).
\bibitem{nazarovbook} Y. V. Nazarov, Y.M. Blanter, Quantum Transport. Introduction to Nanoscience, Cambridge, Cambridge University Press, (2009), Chapter 3.
\bibitem{Keldysh}  L. V. Keldysh, Sov. Phys. JETP, 20, 1018 (1965).
\bibitem{Bekjan} T. Bekjan, Linear Algebra Appl. 390, 321 (2004).
\bibitem{Ando} T. Ando, Linear Algebra Appl. 26, 203-241 (1979). 
\bibitem{nazarov11} Yuli V. Nazarov, Phys. Rev. B 84, 205437 (2011).
\bibitem{ansari3} M.H. Ansari and Y.V. Nazarov, Journal of Experimental and Theoretical Physics 122, 3, pp. 389-401 (2016); 
arXiv:1509.04253.
\bibitem{ansari1}  M. H. Ansari, Yu. V. Nazarov, Phys. Rev. B 91, 104303 (2015), arXiv:1408.3910.
\bibitem{ansari2}  M.H. Ansari and Y.V. Nazarov, Phys. Rev. B 91, 174307 (2015), arXiv:1502.08020. 
\bibitem{ansariST} M.H. Ansari and Y.V. Nazarov, To appear.
\bibitem{ansari4} M.H. Ansari,  arXiv:1605.04620; M.H. Ansari,  arXiv:1512.06288.
\bibitem{schully} M.O. Scully, Phys Rev Lett 104:207701 (2010).
\bibitem{KMS} R. Kubo, J. Phys. Soc. Japan 12, 570 (1957), P. Martin and J. Schwinger, Phys. Rev. 115,
1342 (1959).
\bibitem{cao} J. Ye, K. Sun, Y. Zhao, Y. Yu, C.K. Lee, J. Cao, J. Chem. Phys. 136, 245104 (2012); D. Xu, C. Wang, Y. Zhao, J. Cao,  arXiv:1508.04708.
\bibitem{whaley} S. Hoyer, F. Caruso, S. Montangero, M. Sarovar, T. Calarco, M. B. Plenio and K. B. Whaley, New J. Phys. 16, 045007 (2014).
\bibitem{khoshnegar} M. Khoshnegar, A. Jafari-Salim, M.H. Ansari, A. H. Majedi, New J. Phys. 16 023019 (2014), arXiv:1303.1453. 
\bibitem{kirk} A.P. Kirk, Phys Rev Lett 106, 048703 (2011).
\bibitem{nikonov} D.E. Nikonov, A. Imamoglu, M.O. Scully, Phys Rev B 59, 12212 (1999).
\bibitem{Rahav} S. Rahav, U. Harbola,  S. Mukamel,  Phys. Rev. A 86, 043843 (2012).
\bibitem{schullylasing} S. Ya. Kilin, K. T. Kapale, and M. O. Scully, Phys. Rev. Lett. 100, 173601 (2008).
\bibitem{schullysingle} M. Scully, M. Zubairy, G. Agarwal, and H. Walther, Science 299, 862 (2003).
\bibitem{ansariqp}   M.H. Ansari, Superconductor Science and Technology 28 (4), 045005 (2015) arXiv:1303.1453; M.H. Ansari, F.K. Wilhelm
Physical Review B 84 (23), 235102 (2011) arXiv:1106.4794. 
\bibitem{Kamenev}A. Kamenev, Field Theory of Non-Equilibrium Systems, Cambridge University Press, Cambridge, (2011).
\bibitem{renyi} A. Renyi, Proceedings of the 4th Berkeley Symposium on Mathematics and Statistical Probability, p. 547 (1961).
\bibitem{klish} I. Klich and L. S. Levitov, Phys. Rev. Lett. 102, 100502 (2009).
\bibitem{flindt} K. H. Thomas and C. Flindt, Phys. Rev. B 91, 125406 (2015); Y. Utsumi, Phys. Rev. B 92, 165312 (2015). 
\bibitem{Kindermann} M. Kindermann and S. Pilgram, Phys. Rev. B 69, 155334 (2004).
\bibitem{legett} A. J. Leggett, S. Chakravarty, A. T. Dorsey, M. P. A. Fisher, A. Garg, and W. Zwerger, Rev. Mod. Phys. 59, 1 (1987).



















\end{thebibliography}
\end{document}